\documentclass[aps,prb,twocolumn,nofootinbib,floatfix,bibliography,superscriptaddress]{revtex4-2}
\usepackage{graphics,bm,amsmath,amssymb,natbib,url,epsfig,psfrag,color}
\usepackage{graphicx}
\graphicspath{{./Figs/}{./}} 
\usepackage{braket}
\usepackage[caption=false]{subfig}
\newcommand{\be}{\begin{equation}}
\newcommand{\ee}{\end{equation}}
\newcommand{\bea}{\begin{eqnarray}}
\newcommand{\eea}{\end{eqnarray}}
\def\bal#1\eal{\begin{align}#1\end{align}}

\newcommand{\beq}{\begin{eqnarray}}
\newcommand{\eeq}{\end{eqnarray}}
\usepackage{xcolor}
\usepackage[unicode=true,pdfusetitle,bookmarks=true,bookmarksnumbered=false,bookmarksopen=false,breaklinks=false,pdfborder={0 0 0},pdfborderstyle={},backref=false,colorlinks=true]{hyperref}
\hypersetup{colorlinks,allcolors=blue}

\newcommand{\Tr}{\text{Tr}}
\colorlet{mlc}{orange!70!gray}

\begin{document}

\newcommand{\eran}[1]{{\color{blue}#1}}

\def\bs#1\es{\begin{split}#1\end{split}}	\def\bal#1\eal{\begin{align}#1\end{align}}
\newcommand{\nn}{\nonumber}
\newcommand{\sgn}{\text{sgn}}

\title{Measuring  work in quantum many-body systems using a dynamical ``work agent"}

\author{Cheolhee Han}
\affiliation{Department of Data Information and Physics, Kongju National University, Kongju 32588, Republic of Korea}	
 
\author{Nadav Katz}
\affiliation{Racah Institute of Physics, the Hebrew University of Jerusalem, Jerusalem 91904, Israel}

\author{Eran Sela}
\affiliation{Raymond and Beverly Sackler School of Physics and Astronomy, Tel Aviv University, Tel Aviv 69978, Israel}

\date{\today}
\begin{abstract} 
We consider a generic quantum many-body system  initiated at thermal equilibrium and driven by an external parameter, 
and discuss the prospect for measuring the work done by the varying parameter on the system. While existing methods are based on a full control of the system's Hamiltonian and are thus limited to few-level quantum systems, measuring work in many-body quantum systems remains challenging. Our approach relies on transforming the external parameter into a dynamical ``work agent", for which we consider an harmonic oscillator in a semiclassical coherent state with a large photon number. We define a work generating function which coincides with the standard two-point measurement protocol for work measurement in the limit of a large photon number. While \emph{in principle} it allows to relate the moments of work $\langle W^n \rangle$ to observables of the work agent, we focus on the average work, which is obtained from energy conservation by the change of the energy of the agent, which can be measured using photon number detection. We illustrate this concept on a transmon-microcavity system, 
which displays various quantum coherent effects including Landau-Zener St\"{u}kelberg interference and collapse and revival of Rabi oscillations. We discuss how our setup allows to measure work in a variety of quantum many-body systems.
\end{abstract}
\maketitle

\section{Introduction} 
Work and its statistics have been identified as a promising observable of quantum many body systems driven out of equilibrium. General theoretical studies explored work statistics  in quantum many body systems all the way from the adiabatic limit of slow driving~\cite{PhysRevResearch.2.023377,li2024entanglement} to the quenched limit~\cite{PhysRevLett.109.160601,PhysRevX.4.031029,goold2018role,PhysRevB.90.094304} and near quantum phase transitions~\cite{silva2008statistics,PhysRevE.89.062103,PhysRevResearch.2.033167,fei2020work,rossini2021coherent}. However, measurements of work statistics remain elusive in quantum systems with many degrees of freedom. To be concrete, we consider a quantum many-body system described by a Hamiltonian $H_{ext}(X(t))$ controlled by an external time dependent parameter $X(t)$, and we are interested in measuring the work done by $X(t)$ on the system, where the system is initially at thermal equilibrium. The challenge we propose to overcome  is to allow for work measurements based on the two-point measurement (TPM) protocol~\cite{talkner2007fluctuation}, involving two 
projective measurements of the energy of the system. Although this definition does not capture coherence of energy eigenstates in the initial state, here we are concerned with an initial thermal state. In this case, the TPM scheme still captures coherence generated during the process. The associated work distribution function (WDF) satisfies non-equilibrium fluctuation theorems~\cite{jarzynski1997nonequilibrium,crooks1999entropy,Mandal_2016,PhysRevLett.123.230603}.

Let us first clarify the inadequacy of a direct application of  existing methods for the many-body case. Some of the first measurements of the WDF in the quantum case have been formulated using an ancilla qubit~\cite{PhysRevLett.110.230601,PhysRevLett.110.230602} and demonstrated experimentally using  nuclear magnetic resonance~\cite{batalhao2014experimental}; Soon after, the TPM was formulated as a single positive operator-valued measurement which can be realized by extending the system using an ancilla coordinate~\cite{roncaglia2014work} acting as a ``work meter", see also~\cite{xu2021single}, which is entangled with the system via a quantum operation conditioned by the system's Hamiltonian, before being measured. This protocol had been realized  using cold atoms~\cite{cerisola2017using}, and similar approaches  had been applied in a variety of quantum systems including 
nitrogen-vacancy centers~\cite{PhysRevA.108.042423}, trapped ions~\cite{an2015experimental}, and quantum optics~\cite{de2018experimental,PhysRevA.106.L020201}, to name a few. This allowed to measure the WDF and verify fluctuation theorems in the quantum case, however all of these systems are essentially few-level systems. In a many-body quantum system with many interacting degrees of freedom, it seems hopeless to run the protocols of Refs.~\cite{PhysRevLett.110.230601,PhysRevLett.110.230602,batalhao2014experimental} and directly entangle the many-body system's Hamiltonian with a simple ancillary degree of freedom. Thus, practically these methods are not directly applicable in many-body systems.

As explained in Sec.~\ref{se:review}, our approach is as follows. Our target is a many-body Hamiltonian $H_{ext}(X(t))$ controlled by an external time dependent parameter $X(t)$. We replace the external parameter  $X(t)$ by a dynamical degree of freedom described by an operator $\hat{X}$. Its Hamiltonian and  its initial condition are selected to reproduce $X(t)$ in average with as small fluctuations as possible. For example, a quantum oscillator corresponds to a sinusoidal time dependent parameter, and the small fluctuation limit is achieved using a semiclassical coherent state.  The combined system and ``parameter" evolve under a \emph{time-independent} Hamiltonian $H_{tot}$. Therefore the sum of the average energy of the oscillator alone, and of the rest of the Hamiltonian $H(\hat{X})$, which includes the system and its interaction with the oscillator, are constant in time. Using energy conservation, our elementary statement is that by initiating the oscillator in a semi-classical coherent state with a large number of photons, the average work as defined by the TPM scheme can be measured via the change of the energy stored in the oscillator. 

The idea of treating the work agent dynamically and autonomously is fundamental in quantum thermodynamics and goes beyond the need to measure the WDF. 
In earlier works, to mention a few, it was referred to as a ``weight system"~\cite{skrzypczyk2014work,biswas2022extraction} having just a gravitational potential and interacting via unitary evolution with the system. Many other works consider the work agent as an harmonic oscillator. In a  quantum heat engine this oscillator can be used to alternate the coupling of a system between hot and cold heat reservoirs~\cite{gelbwaser2013work}, again within a fully unitary description with a time-independent Hamiltonian. This oscillator was referred to as a ``quantum flywheel"~\cite{levy2016quantum,PhysRevA.108.L050201} and experimentally realized~\cite{PhysRevLett.123.080602}. The resonator can also serve as a ``quantum battery"~\cite{elouard2015reversible,campaioli2024colloquium}. Since this oscillator evolves with the system in an autonomous way, measurement of its energy fluctuations acts as an embedded quantum work meter~\cite{monsel2018autonomous}, as realized experimentally with a superconducting qubit coupled to a resonator~\cite{cottet2017observing}. 
Based on Ref.~\cite{cohen2012straightforward}, we recently explored the possibility to use the work agent approach to verify the JE~\cite{han2024quantum}.

With the motivation of using the work agent approach to measure the WDF in complicated quantum many-body systems, our key technical advancement described in Sec.~\ref{se:review} is to formulate a generating function  which coincides with that of the TPM scheme (corresponding to an external time-dependent parameter) in the classical limit of infinite number of photons in the coherent state. This generating function allows \emph{in principle} to measure any desired moment of the WDF by inspecting solely the single-coordinate work agent. Adhering to what is practically measured in experiment, the first moment, i.e. the average work, is  nothing but (minus) the change in the number number of photons in the oscillator times $\hbar \omega$, with $\omega$ the oscillator frequency.  

The main purpose of this work, as described in Sec.~\ref{se:outline}, is to  illustrate this concept in a versatile condensed matter system, a transmon-micro-resonator device. This system represents a class of quantum electronic devices controlled by a local time varying gate voltage doing work on the system. The microcavity resonator represents both external time dependent gate voltage, and allows to measure the work done on the system by photon number detection.

In Sec.~\ref{se:results} we exemplify in this system how known quantum interference effects in quantum optics such as Landau-Zener St\"{u}kelberg interference or collapse and revival of Rabi oscillations, are reflected on the average work, and survive in our work-agent drive provided that the number of photons is large enough making the work agent sufficiently classical.

While our focus here is on average work which is practical to measure, we mention in Sec.~\ref{se:JE} that in principle one could measure also the second moment of work, and show that the JE as measured by our work agent will be satisfied in the limit of a large photon number. 

In the extended summary and outlook Sec.~\ref{se:summary}  we discuss the versatility of our proposed setup, by suggesting quantum dot systems as examples of many-body quantum systems, in which thermodynamic proceeses and work measurement can be performed using our setup.

\section{General protocol} 
\label{se:review}
A system is prepared at thermal equilibrium with respect to some Hamiltonian denoted $H_{ext}(X(t_i))$, and then evolves as a closed system under the influence of a time dependent Hamiltonian controlled by an external parameter $H_{ext}(X(t))$ from $t=t_i$ to $t = t_f$.
According to the TPM protocol, the distribution function of the work done on the system by the external parameter is
\bal
\label{eq:2_time}
P(W)=\sum_{n_i,n_f} p(n_i) |\langle n_i|U|n_f\rangle|^2\delta(W-E_{n_f}+E_{n_i}),
\eal
where $\{|n_i \rangle \}$ and $\{E_{n_i} \}$ denote the eigenstates and energies of the initial Hamiltonian $H_{ext}(X(t_i))$, $\{|n_f  \}$ and $\{E_{n_f} \}$ denote the eigenstates and energies of the final Hamiltonian $H_{ext}(X(t_f))$, $U$ is the evolution operator from $t_i$ to $t_f$ under $H_{ext}(X(t))$, and $p(n_i)=e^{-E_{n_i}/T}/Z_{i}$ describes a thermal distribution with respect to the initial Hamiltonian with $Z_i=\sum_{n_i} e^{-E_{n_i}/T}$. It follows that
\bea
h(u)&=&\int dW P(W)e^{-uW} \nonumber \\
&=&\Tr[ e^{-uH_{ext}(X(t_f))} U e^{-(\beta-u)H_{ext}(X(t_i))} U^\dagger]/Z_i.\label{eq:generatingfn}
\eea
$h(u)$ is the generating function of the WDF, allowing to extract the moments from
\bal
\langle W^n \rangle=\int dW W^n P(W)=(-1)^n\frac{d^n}{du^n}h(u)|_{u \to 0}.
\eal

The TPM definition Eq.~(\ref{eq:2_time}) is consistent with fluctuation theorems and satisfies the JE. We now give a protocol to measure it using a dynamical work agent.

We extend the Hilbert space of the system to include the Hilbert space of an  Harmonic oscillator. The total Hamiltonian  is $H_{tot}=H_{\mathcal{S}}+H_{\mathcal{A}}$ with $H_{\mathcal{A}} = \hbar \omega \hat{a}^\dagger \hat{a}$, and the system's Hamiltonian $H_{\mathcal{S}}$ contains the interaction between the system and the agent. Defining a time evolving coherent state $| \alpha(t) \rangle =e^{-i H_{\mathcal{A}} t/\hbar} | \alpha\rangle$ for a decoupled oscillator, where $\hat{a}|\alpha \rangle =\alpha |\alpha \rangle$, we design the system's Hamiltonian such that 
\be
\label{eq:H_condition}
\langle  \alpha(t) |H_{\mathcal{S}} | \alpha(t) \rangle = H_{ext}(X(t)).
\ee

We will consider throughout a linear coupling of the form
\bea
\label{eq:H_0}
H_{ext}(X(t))=H_0+X(t)\mathcal{\hat{O}}, \nonumber \\
H_{\mathcal{S}}=H_0+\underbrace{\hbar \lambda (\hat{a}+\hat{a}^\dagger)}_{\hat{X}} \mathcal{\hat{O}},
\eea
and a sinusoidal variation of the parameter $X(t)=X_0 \cos \omega t$. 
Here $\mathcal{\hat{O}}$  is a system's (dimensionless) operator to which the parameter $X(t)$ couples (see Eq.~(\ref{eq:Ec_EJ}) below), and $\lambda$ is some coupling constant. 

The average in Eq.~(\ref{eq:H_condition}) has a redundancy $\alpha\to \alpha x$, $\lambda \to \lambda x^{-1}$. Denoting the coherent state in terms of its average number of photons, $|\alpha\rangle = |\sqrt{n_{ph}}\rangle$, we  use this freedom to increase the number of photons while weakening the interaction $\lambda$, thus pushing the agent towards the classical regime $n_{ph} \gg 1$. 
For a coherent state $\langle  \alpha |\hat{X} | \alpha \rangle = 2 \hbar \lambda \sqrt{n_{ph}}$ and $\delta \hat{X}=\sqrt{\langle  \alpha |\hat{X}^2 | \alpha \rangle - \langle  \alpha |\hat{X} | \alpha \rangle^2}=\hbar\lambda$. Hence the relative fluctuation of the parameter $\hat{X}(t)$ coupling to the operator $\mathcal{\hat{O}}$ vanishes in the classical limit as  \be
\label{eq:deltaX}
\frac{\delta \hat{X}}{\langle  \alpha |\hat{X} | \alpha \rangle} = \frac{1}{2 \sqrt{n_{ph}}}.
\ee

This motivates us to define a work generating function within the work-agent setup,
\be
\label{eq:h_a}
h_{\mathcal{A}}(u)=\Tr[ e^{-u H_{{\mathcal{S}}}} e^{-iH_{tot}\tau/\hbar} e^{\frac{u}{2}H_{{\mathcal{S}}}}\rho_{initial} e^{\frac{u}{2}H_{{\mathcal{S}}}} e^{i H_{tot} \tau/\hbar}   ],
\ee
with the initial density matrix
\bal
\rho_{initial}=\frac{e^{-\beta H_{ext}(t_i)}}{Z_i}\otimes |\sqrt{n_{ph}}\rangle\langle \sqrt{n_{ph}}|,
\eal
and $\tau=t_f-t_i$.
If we replace the operators $\hat{a}^\dagger,\hat{a}$ by $\alpha^*(t), \alpha(t)$ in all the factors in Eq.~(\ref{eq:h_a}), then $H_\mathcal{S} \to H_{ext}$ and we recover Eq.~(\ref{eq:generatingfn}). This is formally proven in  Appendix~\ref{appendix:h_A}.
We refer to this replacement as a Born-Oppenheimer approximation, which in our case becomes justified when the energy stored in the agent well exceeds the typical work done on the system, $\hbar \omega n_{ph} \gg W_{typical}$. Then, the semiclassical motion of the agent becomes unaffected by the system.
But the Born-Oppenheimer approximation will have corrections for finite $n_{ph}$, that will be explored below. 

Notice that while in Eq.~(\ref{eq:generatingfn}) the initial state commutes with $H_{ext}(t_i)$, in Eq.~(\ref{eq:h_a}) the initial state does not commute with $H_{\mathcal{S}}$. For this reason we separated the operator $e^{u H_{S}}$ into two pieces, see Eq.~(2) in Ref.~\cite{suomela2014moments}. Notice also that when the interaction between the agent and the system vanishes, $\lambda=0$, then $H_{ext}={\rm{const}}$ so that $h(u)={\rm{const}}$, and also $h_{\mathcal{A}}={\rm{const}}$ because $H_{tot}$ commutes with $H_{\mathcal{S}}$ in Eq.~(\ref{eq:h_a}).

With the definition given in Eq.~(\ref{eq:h_a}), we now move to the physical protocol. It starts with the preparation of the initial state $\rho_{initial}$  and continues with its evolution according to $H_{tot}$.

In Ref.~\cite{han2024quantum} it was suggested then to measure the energy of the agent at $t=t_f$ in order to infer the energy change of the system by energy conservation. However the energy measurement of a coherent state suffers from an uncertainty $\propto \sqrt{n_{ph}}$. Minimizing this uncertainty, and also satisfying the Born-Oppenheimer condition, allows access to the WDF only in a limited parameter regime~\cite{han2024quantum}.
Here, using the generating function Eq.~(\ref{eq:h_a}) we formulate a measurement scheme for each work-moment separately. 

\subsection{Moments of the WDF}
\label{se:moments}
First moment of work: By applying a $u$-derivative on Eq.~(\ref{eq:h_a}), thus dropping a factor of $H_S=H_{tot}-H_{\mathcal{A}}$ and noticing that $H_{tot}$ commutes with the evolution operator,  one readily obtains
\bal
\langle W \rangle=\hbar \omega\langle \hat{n}_{ph}(t_f)\rangle-\hbar\omega\langle \hat{n}_{ph}(t_i) \rangle.
\eal
Namely, the average work is obtained, as expected, from an energy measurement of the agent.

Work variance: Higher moments are obtained by applying multiple $u$-derivatives on Eq.~(\ref{eq:h_a}).
For the second moment one obtains (see Appendix~\ref{appendix:moments})   
\bal
\label{eq:2ndmoment}
\frac{\langle W^2 \rangle }{(\hbar \omega)^2}=&\langle \hat{n}_{ph}^2(t_i)\rangle+\langle \hat{n}_{ph}^2(t_f)\rangle-2\text{Re}\langle \hat{n}_{ph}(t_f)\hat{n}_{ph}(t_i)\rangle.
\eal
The first two terms can be extracted by measuring separately the number distribution function at $t=t_i$ and at $t=t_f$. The most complicated term is the two-time correlator in the last term. 

While having a closed form, Eq.~(\ref{eq:2ndmoment}) is not easy to measure experimentally and higher moments become more cumbersome. For this reason, in our implementation below, we focus on the first moment only.

\section{Transmon-microcavity model and proposed experimental protocol} 
\label{se:outline}
\begin{figure}
\centering
\includegraphics[width=.9\columnwidth]{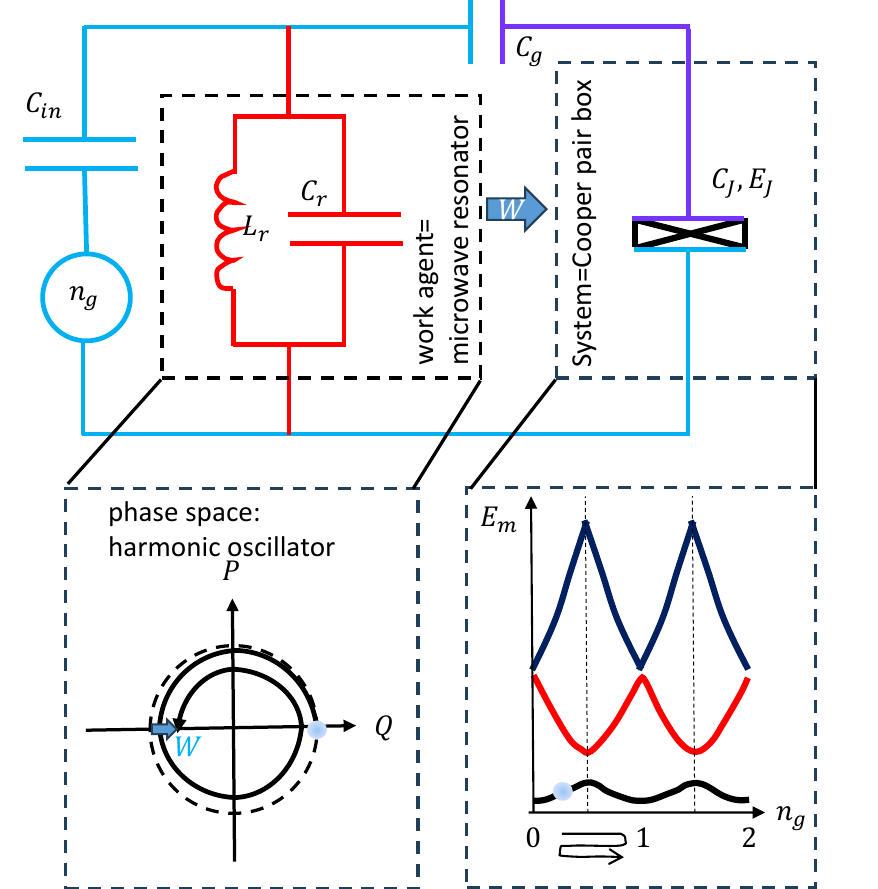}
\caption{Cooper-pair box (CPB) coupled to a micro-resonator as a test bed for work statistics measurement and verification of the Jarzynski's work-fluctuation theorem. The agent is the microwave resonator, which while oscillating exchanges energy with the system  - the CPB. The coordinate of the resonator effectively controls the parameter $n_g$. }
\label{fig:agent_LC}
\end{figure}

We consider  a Cooper-pair box (CPB) (or transmon) coupled to a microwave cavity~\cite{koch2007charge}
\be
\label{eq:Ec_EJ}
H_{tot}=4E_c\hat{n}^2 - V_g \hat{n} -E_J \cos \hat{\varphi} + \hbar \omega \hat{a}^\dagger \hat{a} + \hbar \lambda  \hat{n} (\hat{a} +\hat{a}^\dagger ),
\ee
where $V_g=8E_c n_g$.
As depicted in Fig.~\ref{fig:agent_LC}, the ``system" consists of the CPB, controlled by a charging energy $E_c=\frac{e^2}{2C_\Sigma}$ with $C_\Sigma=C_J 
+C_g$, and a Josephson energy $E_J$, having $n_g$-dependent energy levels $\{ E_m \}$. The system is capacitively coupled to the microwave resonator with  frequency $ \omega=\frac{1}{\sqrt{L_r C_r}}$ with coupling $ \hbar \lambda(t>0)=2 \beta eV_{rms}$, with $\beta=C_g/C_\Sigma$ and $V_{rms}=\sqrt{\hbar \omega/2C_r}$.

Our target is to measure the work done on the system (CPB without resonator) by a time-dependent external gate voltage, as described by 
\be
\label{eq:Hext}
H_{ext}(t)=4E_c\hat{n}^2 - V_g(t) \hat{n} -E_J \cos \hat{\varphi},
\ee
where $V_g=8E_c n_g(t)$ and $n_g(t)=\bar{n}_g+\delta n_g \cos(\omega t)$. Namely, the dimensionless gate voltage oscillates around $\bar{n}_g$ and its initial value is $n_{g,initial}=\bar{n}_g + \delta n_g$; the system is prepared at thermal equilibrium with respect to $H_{ext}(t_i)$.
We refer to this as classical or external driving.

We now employ the microwave resonator as a dynamic work agent, \emph{i.e.} we simulate the external time dependence of Eq.~(\ref{eq:Hext}), by the dynamics of Eq.~(\ref{eq:Ec_EJ}).

\begin{figure}
\centering
\includegraphics[width=.9\columnwidth]{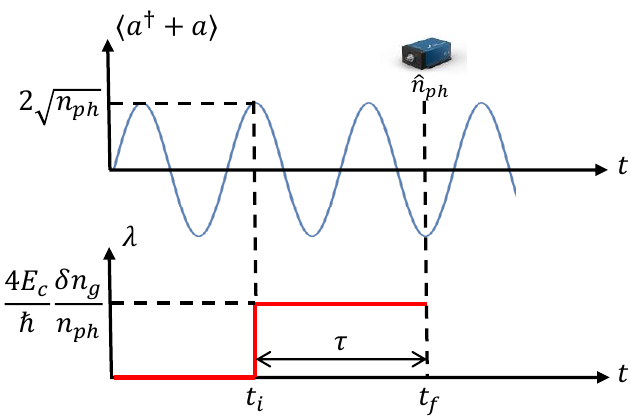}
\caption{Three-stage protocol for work measurement: (i) prepare a coherent state in the resonator, initially decoupled from the system which is in a thermal state. (ii) Turn on suddenly the coupling at time $t_i$. (iii) Measure the photon number in the resonator at time $t_f$. The average work is obtained using Eq.~(\ref{eq:main}). The second moment is obtained using Eq.~(\ref{eq:2ndmoment}).}
\label{fig:protocol}
\end{figure}

The thermodynamic process is illustrated in Fig.~\ref{fig:protocol}. (i) Initially $\lambda(t<0)=0$ and the system is at thermal equilibrium at temperature $T$ with $V_g=8 E_c n_{g,initial}$ where $n_{g,initial}=\bar{n}_g+\delta n_g$. While decoupled from the system, the microwave resonator is prepared in a coherent state $\hat{a}|\alpha \rangle =\alpha |\alpha \rangle$ with $\alpha=\sqrt{n_{ph}}$, with $\langle \hat{a}  \rangle_t =\sqrt{n_{ph}} e^{-i \omega t}$ and $\langle \hat{a}^\dagger  \rangle_t =\sqrt{n_{ph}} e^{i \omega t}$,   such that
\be
\label{eq:deltan_g}
   \delta n_g = - \frac{ \hbar \lambda}{4E_c} \sqrt{n_{ph}}, 
\ee
where $n_{ph}=\langle \hat{a}^\dagger \hat{a} \rangle_{t=0}$ is the initial average number of photons in the resonator.  (ii) Then,
the coupling $\lambda$ is turned on at time $t_i=0$ which is the beginning of the process, where we also suddenly change the gate voltage from $n_{g,initial}$ to $n_{g,final}=\bar{n}_{g}$. 
So, 
what really happens at the initial time is that the external parameter (having no fluctuations) is replaced by the work agent which coincides with the parameter on average but introduces fluctuations. The central point is that in the limit of a large number of photons these fluctuations vanish, see Eq.~(\ref{eq:deltaX}).
(iii) The process ends at time $t_f=\tau = \frac{\pi (2m+1)}{\omega}$ with some integer $m$. The photon number is then measured. 

Although a more complicated analysis is required to measure higher moments of work (see Sec.~\ref{se:moments}), we focus on the average work.  By energy conservation, the work done by the resonator on the transmon is~\cite{cottet2017observing}
\be
\label{eq:main}
\langle W \rangle=\hbar \omega (\langle \hat{a}^\dagger \hat{a} \rangle_{t=0}-\langle \hat{a}^\dagger \hat{a}\rangle_{t=\tau}). 
\ee

When do we expect  work statistics extracted from the above protocol to coincide with that of a classical drive $H_{ext}(t)$ using the TPM scheme?  As we argued in Sec.~\ref{se:review}, and demonstrate  numerically below in various regimes, for a large number of photons,  $n_{ph} \gg 1$, the work agent becomes classical, and therefore the total Hamiltonian $H_{tot}$ depending on the operator $\hat{a}, \hat{a}^\dagger$ becomes the Hamiltonian $H_{ext}$ controlled by an external parameter; Yet due to energy conservation any change in the internal energy of the system is accounted for by an opposite change in the energy of the work agent, which is our suggested work measurement.  

In other words, when $n_{ph} \gg 1$  the typical work 
becomes smaller than the energy stored in the oscillator, 
\be
\label{eq:BO}
W_{typical} \ll \hbar \omega n_{ph},
\ee
so the system is a small perturbation on the agent's dynamics.
This condition is called the Born-Oppenheimer condition~\cite{han2024quantum}. It is then legitimate to replace $\hat{a}$ and $\hat{a}^\dagger$ in Eq.~(\ref{eq:Ec_EJ}) by their quantum averages.

\section{Results}
\label{se:results}
The CPB has various regimes depending on the ratios $E_J/E_c$, $\omega/E_J$, and $\delta n_g$. Our result that work statistics can be accurately extracted using the agent for a large photon number applies in general. Let us discuss two regimes (see Fig.~\ref{fig:interference_coherent_state}). 

\begin{figure}
    \centering
    \includegraphics[width=0.95\linewidth]{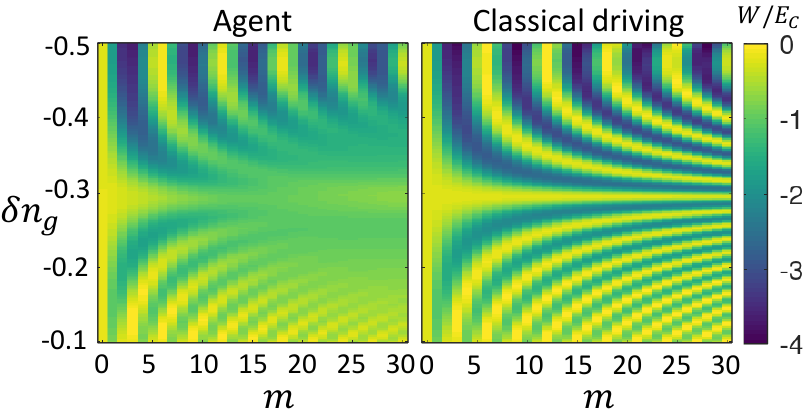}
    \caption{
    Average work obtained using the agent protocol (left) or  external driving (right) as a function of $\delta n_g$ and number of periods $m$, where the total time of the process is $\tau=\frac{\pi(2m+1)}{\omega}$ with integer $m$. The oscillations stem from Landau-Zener-St\"{u}kelberg interference. Here we use $E_c=\hbar \omega$, $E_J=0.4\hbar \omega$, $T=0.25\hbar \omega$, and $n_{ph}=100$. Qualitatively, work is similar to the classical driving. 
    }
    \label{fig:2d_plot}
\end{figure}

(i) Landau-Zener regime, in which transitions between energy levels are governed by individual avoided crossings. This occurs for $E_J \ll E_c$ and $\delta n_g \gtrsim 1/2$. An example is in Fig.~\ref{fig:2d_plot}. Here the frequency $\omega$ dictates the Landau-Zener transition probability. At the same time we use the agent to measure work, so $\hbar \omega$ is the energy resolution. Nonetheless, our method applies even in the large frequency regime, although this has a price - a type of superresolution.

(ii) Rabi oscillations regime: $E_J \gg E_c$ (transmon limit) or $\delta n_g \ll 1$ where $E_m$ are almost constant. In this regime, it is particularly interesting to consider the resonant case between the agent and the system.

\subsection{Landau-Zener regime and LZS interference}

In the left panel of Fig.~\ref{fig:2d_plot} we show a simulation of our  protocol for $T=0.25\hbar \omega, E_J=0.4\hbar \omega, E_c=8\hbar \omega$, $\bar{n}_g=0.5$ and 
photon number $n_{ph}=100$. The resulting $\langle W \rangle$ is plotted both versus $\delta n_g$ (controlled by $\lambda$) and $m$ and displays Landau-Zener-St\"{u}kelberg (LZS) interference, as discussed next. The coupling $\lambda$ varies with $\delta n_g$ as determined from Eq.~(\ref{eq:deltan_g}), and its maximum value is $\lambda=0.05  \omega$. Our protocol with the work agent is compared with the TPM results in the right panel of Fig.~\ref{fig:2d_plot}. 

 The LZS oscillations in the work are an example of a coherent effect that is created during the thermodynamic  process (which starts with a thermal state). Fig.~\ref{fig:2d_plot} emphasizes the fact that these oscillations are maintained when the process is driven by the agent despite its finite fluctuation which are expected to reduce quantum oscillations, as quantified below, see Eq.~(\ref{eq:quantum_coherence}).

Let us discuss the LZS oscillations.
When an avoided crossing is crossed multiple times due to the oscillating parameter $n_g(t)$, since in each event there is a finite  amplitude for an adiabatic passage or for a Landau-Zener transition, there appear interference effects as seen in Fig.~\ref{fig:2d_plot}. But when the parameter $n_g(t)$ is replaced by the work agent coordinate $\propto (\hat{a}+\hat{a}^\dagger )$, the quantum fluctuations of the latter which scales as $\sqrt{n_{ph}}$, can wash out the relative phases. The key question then is: under what conditions will the dynamic work agent coherently simulate the process $H_{ext}(t)$?
As we  show  in Appendix~\ref{se:coherence},  the work agent   faithfully captures the LZS  interference  provided that
\be
\label{eq:quantum_coherence}
\lambda \ll  \omega~~~\to~\rm{{LZS~interference}}.
\ee
Crucially, in the $n_{ph} \to \infty$ limit, keeping $\delta n_g$ constant we can see from Eq.~(\ref{eq:deltan_g}) that this condition will be satisfied. 

\begin{figure}
\centering
\includegraphics[width=0.8\columnwidth]{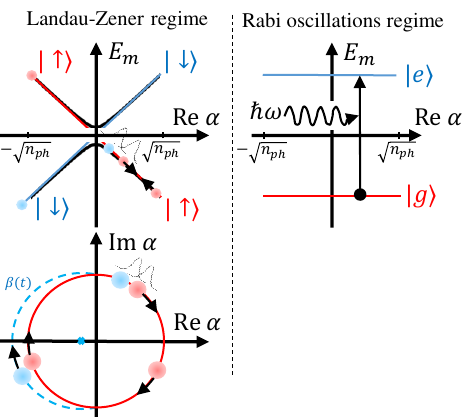}
\caption{Landau-Zener regime versus the Rabi oscillations regime. The typical work is $W_{typical} \sim E_c$ in the Landau-Zener regime (for $\delta n_g \sim 1/2$) and $W_{typical} \sim E_J$ in the Rabi oscillations regime. Bottom panel: In the Landau Zener regime, LZS interference is suppressed due to coherent state overlap. The TLS energy splitting is controlled by the coherent state amplitude ${\rm{Re}}~ \alpha$. The dynamics of the coherent state depends on the TLS state. }
\label{fig:interference_coherent_state}
\end{figure}

Let us illustrate the key argument for this inequality. For subsequent analysis, let us consider the lowest pair of levels at $n_g=1/2$ and define a Pauli matrix $\sigma_z=1-2\hat{n}$. Eq.~(\ref{eq:Ec_EJ}) becomes a two-level system coupled to an harmonic oscillator,
\be
\label{eq:TLS}
H_{TLS}=\hbar  \lambda \frac{1-\sigma_z}{2} (\hat{a}+\hat{a}^\dagger )- \frac{E_J}{2}\sigma_x + \hbar \omega \hat{a}^\dagger \hat{a} .
\ee
The corresponding external Hamiltonian at $\bar{n}_g=1/2$ is
\be
\label{eq:H_TLS_ext}
H_{TLS,ext}=
-8 E_c \delta n_g \cos(\omega t) \frac{1-\sigma_z}{2} -  \frac{E_J}{2}\sigma_x. 
\ee
In order to elucidate the physical origin of the quantum~coherence condition, consider $n_g=1/2$ and let $E_J \ll E_c$ so that energy eigenstates are approximately $\sigma_z-$eigenstates except near the gap closing point (see left panel of Fig.~\ref{fig:interference_coherent_state}). The TLS Hamiltonian is controlled by ${\rm{Re}}( \alpha)$ and the transition between $\sigma_z=+1=\uparrow$  and $\sigma_z=-1=\downarrow$  occurs upon approaching  ${\rm{Re}} ( \alpha) = 0$. For $\sigma_z=\uparrow$ the coherent state follows circular trajectories in the complex-$\alpha$ plane centered at $\alpha=0$, $|\alpha(t)\rangle=|\sqrt{n_{ph}} e^{-i \omega t}\rangle$. For  $\sigma_z=\downarrow$ the Harmonic oscillator is shifted and circular motion is centered at $\alpha=-\frac{\lambda}{ \omega}$.

Consider an initial state 
\be
|\Psi (t=0) \rangle = | \uparrow \rangle \otimes | \sqrt{n_{ph}} \rangle.
\ee
Before it reaches the first Landau-Zener transition, it evolves as $|\Psi (t) \rangle = | \uparrow \rangle \otimes | \sqrt{n_{ph}} e^{-i \omega t}\rangle$. Just after the first Landau-Zener transition we have an entangled system-agent state
\be
|\Psi (t) \rangle =\sqrt{P_{LZ}} | \uparrow \rangle \otimes | \sqrt{n_{ph}} e^{-i \omega t}\rangle + \sqrt{1-P_{LZ}} | \downarrow \rangle \otimes | \beta(t)\rangle,
\ee
where $\beta(t)$ is denoted in Fig.~\ref{fig:interference_coherent_state} and given by $\beta(t)=-\frac{\lambda}{ \omega} +i e^{-i \omega t}\left(\frac{\lambda}{ \omega}+i \sqrt{n_{ph}}\right)$, and  $P_{LZ}$ is the Landau-Zener probability.  The $\sigma_z=\downarrow$ blob ($\beta(t)$) reaches the next Landau Zener crossing point ${\rm{Re}}\beta(t)=0$ at a time $\Delta t$ after the  $\sigma_z=\uparrow$ blob, with $\Delta t
\cong \frac{2 \lambda}{ \omega^2 \sqrt{n_{ph}}}$. For times larger than the second Landau-Zener crossing for both blobs, the  probability amplitude to find the TLS in the ground state is
\be
\langle \uparrow | \Psi (t)\rangle = P_{LZ} | \sqrt{n_{ph}} e^{-i \omega t} \rangle+(1- P_{LZ}) | \sqrt{n_{ph}} e^{-i \omega (t+\Delta t)}   \rangle.
\ee
Therefore the corresponding probability $|\langle \uparrow | \Psi (t)\rangle|^2$ contains the interference term $2 P_{LZ}(1- P_{LZ}) {\rm{Re}} \langle  \sqrt{n_{ph}} | \sqrt{n_{ph}}  e^{i \omega \Delta t} \rangle$.  The overlap between the two coherent states, using the formula $\langle \alpha | \beta \rangle = e^{-\frac{1}{2}|\alpha-\beta|^2}$,  is precisely the limiting factor for coherence, and yields Eq.~(\ref{eq:quantum_coherence}).

\subsection{Resonant work agent and Rabi oscillations} 
\label{se:rabi}
Consider now the  regime
\be
\label{eq:resonant_cond}
\hbar \omega \approx E_J \gg \hbar \lambda \sqrt{n_{ph}},~~~~E_J \ll E_c.
\ee
Here single photons from the resonator can be absorbed or emitted. We further assume that this occurs at the Rabi frequency $\lambda \sqrt{n_{ph}}/2=2E_c \delta n_g$, which is much smaller than $\omega$. In the eigenbasis of the $E_J$-term, the external Hamiltonian maps to the Jaynes–Cummings model (see Appendix~\ref{appendix:JC})
\bea
H_{JC,ext}&=& \frac{E_J}{2}(|e\rangle \langle e| -|g\rangle \langle g| ) ) \nonumber \\
&-&\frac{\hbar \lambda \sqrt{n_{ph}}}{2} (e^{i \omega t}|g\rangle \langle e| +e^{-i \omega t}|e\rangle \langle g| ).
\eea
The work performed by the drive (work agent) on the system coherently oscillates at the Rabi frequency. This is shown in Fig.~\ref{fig:resonance}. Comparing the classical driving to the agent drive we see various effects familiar from quantum optics known as collapse and revival~\cite{meunier2005rabi}:
\begin{figure}
\centering
\includegraphics[width=.9\columnwidth]{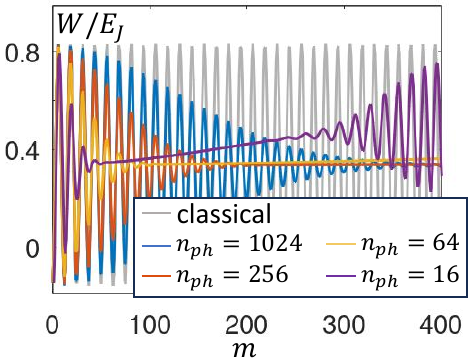}
\caption{Work extracted from the work agent in the resonant regime. Comparison between agent with different $n_{ph}$ and classical driving. Here we set $E_c=E_J=\hbar\omega$, $T=0.01\hbar\omega$, and $\delta n_g=0.02$. As $n_{ph}$ increases, the decay of the Rabi oscillations is postponed to longer times. For the smallest $n_{ph}$ a revival of the Rabi oscillations is visible.}
\label{fig:resonance}
\end{figure}
Here, the coherent oscillations are simply the Rabi oscillations, and in the agent drive the decay occurs  due to the dispersion of the Rabi frequency $\sim \lambda/\hbar$. The decay of Rabi oscillations due to the dispersion of the Rabi frequency, for example of a coherent state, is   well known experimentally~\cite{meunier2005rabi}, to occur after $\sim \sqrt{n_{ph}}$ Rabi oscillations. 
Interestingly, the coherent oscillations partially revive after $\sim n_{ph}$ oscillations~\cite{meunier2005rabi}. As we can see by comparing the classical drive and agent drive in Fig.~\ref{fig:resonance}, that the two progressively agree up to longer times upon increasing the photon number in the resonator.   This example again demonstrates  the relation
between the autonomous and semi-classical definitions of work~\cite{dann2023unification}.

Note that in this resonant regime the typical work  is  $W_{typical} \sim \hbar \omega$, requiring to measure a unit change of the number of photons in average using Eq.~(\ref{eq:main}).

\section{Testing the Jarzynski equality and scaling with $n_{ph}$}
\label{se:JE}

We now return to the generating function of the WDF introduced in Sec.~\ref{se:review}, which in principle allows to measure higher moments of the WDF. Despite the fact that the expression for the second moment is hard to be measured experimentally, it is straightforward to compute it. Like the first moment, also the second moment of the WDF will tend to that of the TPM scheme only in the limit of $n_{ph} \to \infty$. The goal of this section is to confirm the validity of the expression Eq.~(\ref{eq:2ndmoment}) via a test of the  JE in the limit $n_{ph} \to \infty$. 

\begin{figure}
    \centering
    \includegraphics[width=0.99\linewidth]{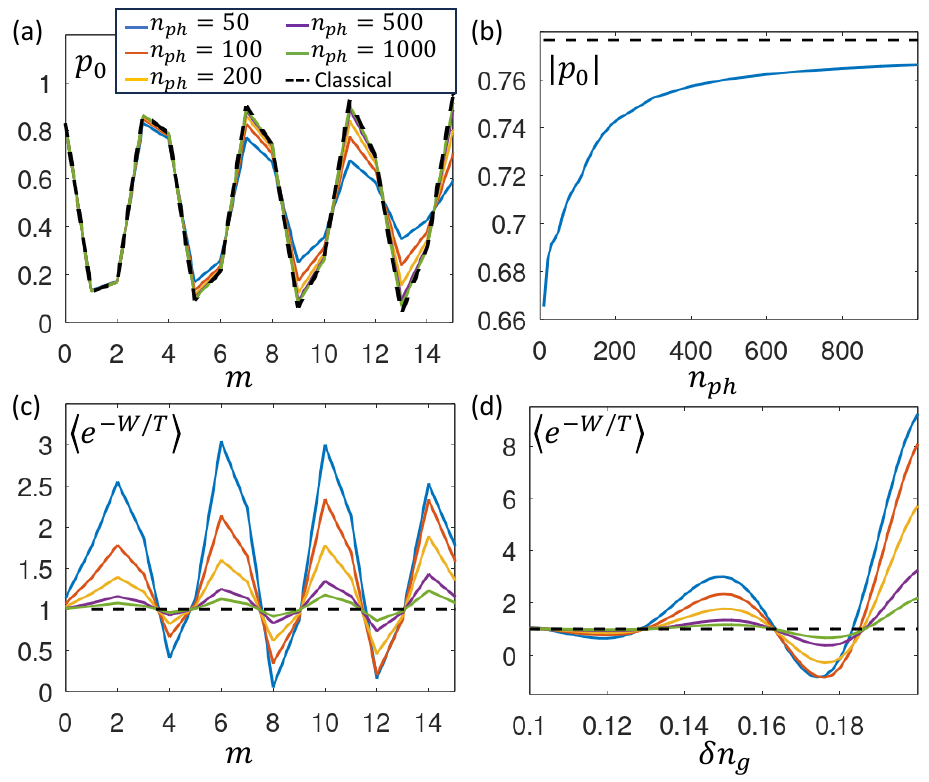}
    \caption{LZS interference pattern for probability $p_0$ and $\langle e^{-W/T}\rangle$. Here we set $E_c=\hbar\omega$, $T=0.25\hbar\omega$, $E_J=0.4\hbar\omega$. (a) $p_0$ as a function of $m$, for various $n_{ph}$ (=50, 100, 200, 500, 1000) and classical driving. Here we fix $\delta n_g=-0.15$ and vary the number of photons while simultaneously adjusting $\lambda$ according to Eq.~(\ref{eq:deltan_g}). As $n_{ph}$ increases, the interference pattern approaches that of classical driving. (b) The amplitude of the interference as a function of $n_{ph}$ (blue line) and classical driving (black dashed). (c) $\langle e^{-W/T}\rangle$ as a function of $m$ with $\delta n_g=0.15$. (d) $\langle e^{-W/T}\rangle$ versus $\delta n_g$ for $m=5$.}
    \label{fig:dephasing}
\end{figure}

The JE says that, under the external protocol Eq.~(\ref{eq:Hext}), the work as defined by the TPM  protocol identically satisfies $\langle e^{-(W-\Delta F)/T} \rangle=1$. $\Delta F$ is the free energy change between the initial and final Hamiltonians, and in our examples we will always consider $\Delta F=0$ (in the transmon this is achieved by symmetric oscillations around $n_g=1/2$).  As we now explicitly demonstrate, the value of $\langle e^{-W/T} \rangle$ extracted using our protocol via the agent, approaches unity as the number of photons in the coherent state increases, see Fig.~\ref{fig:dephasing}.

Let us focus on the TLS. The target WDF is obtained from the external drive in Eq.~(\ref{eq:Hext}). For this simple model the WDF has three delta-function peaks, $P_{TLS}(W)=p_0 \delta(W-W_0)+p_1 \delta(W-W_1)+p_2 \delta(W-W_2)$. Here, $p_0$ describes a  transition  from the ground state to the excited state, $p_1$ describes an adiabatic process where the TLS stays in the initial state (either ground or excited state), and $p_2$ describes the transition probability  from the excited state to the ground state. Each one of these probabilities is a product of the probability of the initial state times the transition probability~\cite{han2024quantum}. The eigenenergies of Eq.~(\ref{eq:H_TLS_ext}) are $E_{\pm}=-4E_c \delta n_g \pm \sqrt{(4E_c \delta n_g)^2+(E_J/2)^2}$, hence the energy differences of these three processes upon going from  $+\delta n_g$ to $-\delta n_g$ (with $\delta n_g<0$) are
\bea
W_0&=&E_+(\delta n_g)-E_-(\delta n_g), \nonumber \\
W_1&=&E_-(\delta n_g)-E_-(\delta n_g)=E_+(\delta n_g)-E_+(\delta n_g), \nonumber \\
W_2&=&E_-(\delta n_g)-E_+(\delta n_g).
\eea
The probability $p_0$, as an example, is seen in Fig.~\ref{fig:dephasing}(a) to nicely compare with the corresponding probability extracted from the dynamical work agent protocol. The approach of the LZS amplitude (the amplitude of the first oscillation)
towards the external driving limit is shown in Fig.~\ref{fig:dephasing}(b) versus $n_{ph}$. 

This comparison between the external drive protocol and the work agent protocol can be tested more strictly via the JE. The WDF $P_{TLS}(W)$ consisting of 3-delta function peaks, by virtue of being defined using the two-time measurement protocol,  satisfies the  JE. We would like now to test the agent protocol, but the JE probes all moments of the WDF. Hence we will make an approximation.
A property of a 3-delta function peak WDF is that it can be fully reconstructed  from the first two moments. Indeed from the pair of quantities
\bea
\langle W \rangle &=& p_0 W_0+p_1 W_1+p_2 W_2, \nonumber \\
\langle W^2 \rangle &=& p_0 W_0^2+p_1 W_1^2+p_2 W_2^2,
\eea
together with $p_0+p_1+p_2=1$, all three peaks can be reconstructed. Let us assume that this reconstruction applies for the full WDF of the agent. This is of course an approximation since the delta function peaks are smeared and shifted due to the agent response~\cite{han2024quantum}, but this approximation becomes exact in the large limit $n_{ph}$.

In Fig.~\ref{fig:dephasing}(c,d) we first evaluate the first and second moments of work. While the first moment is obtained from the average photon number, Eq.~(\ref{eq:main}), the second moment requires photon correlations of the agent, using Eq.~(\ref{eq:2ndmoment}). 
We compute these correlations in our simulation. Then, using the above reconstruction of the WDF, we evaluate $\langle e^{-W/T} \rangle$. This particular combination of all the moments should not display any interference  and is fixed to unity, see dashed lines. The observed oscillations reflect an error of the agent measurement protocol. However, this error diminishes upon increasing the photon number in the agent's coherent state, in accordance with the Born-Oppenheimer condition.

\section{Summary and outlook} 
\label{se:summary}
We developed the concept of the dynamical work agent to measure work statistics. We generalized our recent approach~\cite{han2024quantum}, in which work statistics is obtained solely from energy measurements of the agent, which is limited by energy uncertainty, while here we designed particular observables of the agent whose quantum average coincides with desired moments of the work distribution. 

Using this approach, we theoretically demonstrated the possibility of measuring the average work  on the GHz scale by focusing on a Cooper pair box (transmon) coupled to a microwave resonator, based on photon number detection. In this system we captured quantum coherent effects such as Landau-Zener-St\"{u}kelberg interference or Rabi oscillations and their collapse and revival.  The particular model Hamiltonian we studied, was selected as a proof of concept. A key motivation for an experimental demonstration of our protocol   is to open the door to explore the work statistics in nontrivial quantum many-body systems.

\begin{figure}
\centering
\includegraphics[width=.9\columnwidth]{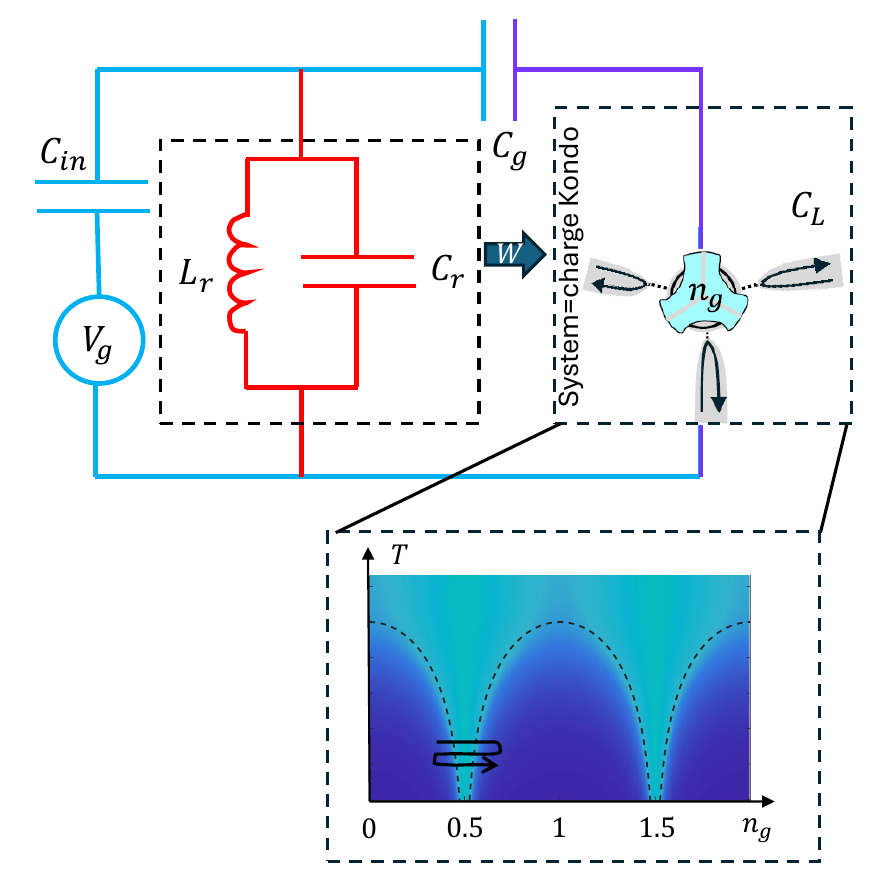}
\caption{Hybrid quantum dot-microresonator system: we replace the qubit Josephson junction by a metallic quantum dot  connected to $M$ (here $M=3$) leads realizing the $M$-channel Kondo critical state. Sweeping the gate voltage $n_g$ on the GHz regime allows to probe work statistics with nontrivial Kibble-Zurek scaling regimes using Eq.~(\ref{eq:main}). }
\label{fig:agent_LC_Kondo}
\end{figure}
\subsection{Challange of measuring work statistics in quantum dots}
One could replace the transmon by a quantum dot, and this allows one to measure the work in mesoscopic quantum dot systems under a time dependent gate voltage, which so far had been limited to the classical in-coherent regime. To illustrate the difficulty, consider the usual setup of a quantum dot tunnel coupled to leads, where the dot's energy level is controlled by a time dependent gate voltage.  The WDF had been measured using continuous charge detection~\cite{saira2012test,koski2013distribution,PhysRevB.93.035425,hofmann2017heat,barker2022experimental}, confirming the Crooks relation~\cite{saira2012test}, JE~\cite{hofmann2017heat} and nonequilibrium fluctuation-dissipation-relations~\cite{barker2022experimental}. Another approach to measure work in quantum dots is via
calorimetric measurements~\cite{pekola2013calorimetric}.
However, quantum many-body effects can not be included in these existing work measurement schemes due to two reasons. First, the continuous charge measurement disrupts the quantum superposition created during the process. Second, to capture quantum effects, it is needed to confine the entire thermodynamic process to time scales shorter than the coherence time, and for quantum dot systems cooled down to $\sim 10 $mK, this requires a $\sim 10^{9}$Hz resolution, which is far beyond the time resolution of current schemes based on charge detection.

One possibility is to consider, as in Fig.~\ref{fig:agent_LC_Kondo}, a hybrid quantum dot-microresonator system. Experiments demonstrating the coupling of double quantum dots to microwave resonators have been demonstrated~\cite{frey2012dipole,Petersson_2012,PhysRevLett.107.256804,PhysRevLett.110.066802} including in the strong coupling limit~\cite{
Bruhat_2018,Mi_2017,PhysRevX.7.011030}. 

\subsection{Coupling a microcavity to a quantum critical system}

Consider a quantum dot displaying nontrivial many-body quantum criticality. In Fig.~\ref{fig:agent_LC_Kondo} we display a metallic quantum dot realizing the charge Kondo effect~\cite{iftikhar2015two,*mitchell2016universality,iftikhar2018tunable,piquard2023observing}, which becomes, for the case of more than one lead, a quantum critical point~\cite{PhysRevB.69.115316} with nontrivial properties~\cite{PhysRevLett.116.157202,han2022fractional,pouse2023quantum,*karki2023z,PhysRevLett.130.136201}. In  this system, the gate voltage $n_g$ acts as a tuning parameter of the quantum critical point. Recently, exact results appeared for the work statistics of this family of quantum critical points~\cite{ma2025quantum}.  The nontrivial regimes enter for time scales $\tau \sim \frac{\hbar}{k_B T}  \sim 10^{-9} $ sec. Thus, combining the quantum dot gate with a microwave resonator could be a promising route to explore novel regimes in stochastic thermodynamics in such many-body quantum systems.

\begin{acknowledgments}
We thank discussions with Josh Folk, Klaus Ensslin, Frederic Pierre, and  Doron Cohen. ES acknowledges support from the European Research Council (ERC) under the European Union Horizon 2020 research and innovation program under grant agreement No.~951541. CH acknowledges support by the research grant of Kongju National University in 2024.
\end{acknowledgments}

\begin{appendix}
\begin{widetext}
\section{Relation between $h(u)$ and $h_\mathcal{A}(u)$}
\label{appendix:h_A}
Here we compute $h_A(u)$ and prove it approaches $h(u)$ as $n_{ph}\to\infty$. First consider $h_\mathcal{A}$, 
\be
\label{eq:h_A_appendix}
h_{\mathcal{A}}(u)=\Tr[ e^{-u H_{{\mathcal{S}}}} e^{-iH_{tot}\tau/\hbar} e^{\frac{u}{2}H_{{\mathcal{S}}}}\frac{e^{-\beta H_{ext}(t_i)}}{Z_i}\otimes |\sqrt{n_{ph}}\rangle\langle \sqrt{n_{ph}}| e^{\frac{u}{2}H_{{\mathcal{S}}}} e^{i H_{tot} \tau/\hbar}   ].
\ee
Using Eq.~(\ref{eq:H_0}), the time evolution operator for the system and agent can be expanded as a power series of $\lambda$,
\bal
U_A=e^{-i(H_0+H_\mathcal{A})\tau/\hbar}\Big(1-\frac{i\lambda}{\hbar}\int_0^\tau dt \hat{\mathcal{O}}(t)(\hat{a}e^{-i\omega t}+\hat{a}^\dagger e^{i\omega t}) +\cdots+\frac{1}{n!}\Big(-\frac{i\lambda}{\hbar}\Big)^n\prod_{i=1}^{n}\int_0^\tau dt_i \prod_{j=1}^n \hat{\mathcal{O}}(t_j)(\hat{a} e^{-i\omega t_j}+\hat{a}^\dagger e^{i\omega t_j}) +\cdots\Bigg).
\eal
Similarly,
\bal
e^{-uH_S}=e^{-u H_0}\Bigg(1-\lambda\int_0^u du_1 \hat{O}(u_1)(\hat{a}+\hat{a}^\dagger)(u_1)+\cdots+\frac{(-\lambda)^n}{n!}(\hat{a}+\hat{a}^\dagger)^n\prod_{i=1}^n \int_0^u du_i \prod_{j=1}^n \hat{O}(u_j) +\cdots\Bigg).
\eal
Now we compute the partial trace of agent part in the $h_A(u)$. The agent part of the initial density matrix is the coherent state, to compute the expectation value of $\hat{a}$ and $\hat{a}^\dagger$ operators, we need to arrange the operators, placing $\hat{a}^\dagger$ to the left of $\hat{a}$. In this process commutators of $\hat{a}$ and $\hat{a}^\dagger$ appear. First we neglect the commutators, and then the result is simply obtained by replacing $\hat{a}, \hat{a}^\dagger \to \sqrt{n_{ph}}$, which reproduces $h(u)$ from Eq.~\eqref{eq:H_TLS_ext}.

Next we consider the commutators. Consider one term which has $\lambda^n$ power. There must be the same order term without commutator, whose contribution is of order $\lambda^n n_{ph}^{n/2}$. Next we focus on the term which contains only one commutator between $b$ and $b^\dagger$. Then its contribution is order of $\lambda^n n_{ph}^{n/2-1}$. Then the ratio between two terms are $1/n_{ph}$. If we have $m$ commutators, then its contribution is order of $\lambda^n n_{ph}^{n/2-m}$, whose ratio is $1/n_{ph}^m$. 
Therefore in the limit $n_{ph}\to \infty$, the error becomes neglected.

This proves that the work and its moments defined using TPM is the same as the energy difference of the energy in the limit of $n_{ph}\to\infty$.


\section{First moments of WDF from correlations of  $a^\dagger a$  }
\label{appendix:moments}
Here we derive general expressions for  moments of the work distribution function in terms of multi-point correlators of the photon number $\hat{n}_{ph}(t)=e^{i H_{tot} t}\hat{n}_{ph}e^{-i H_{tot} t}$ (in this appendix $\hbar=1$). We first define the system's Hamiltonian 
\be
H_{\mathcal{S}}=H_{tot}-H_{\mathcal{A}},~~~H_{\mathcal{A}}=\omega \hat{n}_{ph},
\ee
and use it to define a new generating function corresponding to Eq.~\eqref{eq:generatingfn},
\bal
h_{\mathcal{A}}(u)=\Tr[e^{\frac{u}{2}H_{S}}\rho e^{\frac{u}{2}H_{S}} e^{i H_{tot} t_f} e^{-u H_{S}} e^{-iH_{tot}t_f} ].
\eal
Following the TPM protocol, the transformation with respect to $u$ gives the difference of the system's energy, as measured by $H_{\mathcal{S}}$, between initial state $\rho$ and final state $e^{-i H_{tot} t_f} \rho e^{i H_{tot} t_f}$.
Explicitly,
\bal
h_{\mathcal{A}}(u)=&\Tr[e^{\frac{u}{2}(H_{tot}-\omega\hat{n}_{ph})}\rho e^{\frac{u}{2}(H_{tot}-\omega \hat{n}_{ph})} e^{iH_{tot} t_f} e^{-u (H_{tot}-\omega \hat{n}_{ph})} e^{-iH_{tot}t_f}].
\eal
To obtain the average work, we take the derivative with respect to $u$ at $u\to0$, yielding
\bal
\langle W \rangle =\omega\langle \hat{n}_{ph}(t_f)\rangle-\omega\langle \hat{n}_{ph}(t_i) \rangle.
\eal

To obtain higher moments, we use the Zassenhaus formula, giving
\bal
e^{\frac{u}{2}(H_{tot}-\omega\hat{n}_{ph})}=&e^{\frac{u}{2}H_{tot}}e^{-\frac{u}{2}\omega\hat{n}_{ph}}e^{-\frac{u^2}{8}[H_{tot},\omega\hat{n}_{ph}]}\times e^{\mathcal{O}(u^3)}\nn\\
=&e^{-\frac{u^2}{8}[\omega\hat{n}_{ph},H_{tot}]}e^{-\frac{u}{2}\omega\hat{n}_{ph}}e^{\frac{u}{2}H_{tot}}\times e^{\mathcal{O}(u^3)}.
\eal
Plugging this into the generating function gives up to second order in $u$
\bal
h_{\mathcal{A}}(u)=&\Tr[e^{\frac{u}{2}H_{tot}}e^{-\frac{u}{2}\omega\hat{n}_{ph}}e^{\frac{u^2}{8}[H_{tot},\omega\hat{n}_{ph}]}\rho e^{-\frac{u^2}{8}[\omega\hat{n}_{ph},H_{tot}]} \nn\\
&\times e^{-\frac{u}{2}\omega \hat{n}_{ph}}e^{\frac{u}{2}H_{tot}}e^{iH_{tot} t_f} e^{-\frac{u}{2} H_{tot}}e^{\frac{u}{2}\omega \hat{n}_{ph}}e^{\frac{u}{2}\omega \hat{n}_{ph}}e^{-\frac{u}{2} H_{tot}} e^{-iH_{tot}t_f}]\nn\\
=&\Tr[e^{-\frac{u}{2}\omega\hat{n}_{ph}}e^{\frac{u^2}{8}[H_{tot},\omega\hat{n}_{ph}]}\rho  e^{-\frac{u^2}{8}[\omega\hat{n}_{ph},H_{tot}]} e^{-\frac{u}{2}\omega \hat{n}_{ph}}e^{iH_{tot} t_f} e^{\frac{u}{2}\omega \hat{n}_{ph}}e^{\frac{u}{2}\omega \hat{n}_{ph}} e^{-iH_{tot}t_f}].
\eal
By taking the second derivative with respect to $u$ at $u\to 0$, we obtain the variance of the work
\bal
\langle W^2 \rangle=&\omega^2\Tr[\rho \hat{n}_{ph}^2]+\omega^2\Tr[\rho e^{iH_{tot}t_f}\hat{n}_{ph}^2 e^{-iH_{tot}t_f}]-\omega^2\Tr[\rho e^{iH_{tot}t_f}\hat{n}_{ph}e^{-iH_{tot}t_f}\hat{n}_{ph}]-\omega^2\Tr[\rho\hat{n}_{ph} e^{iH_{tot}t_f}\hat{n}_{ph}e^{-iH_{tot}t_f}]\nn\\
=&\omega^2\langle \hat{n}_{ph}^2(t_i)\rangle+\omega^2\langle \hat{n}_{ph}^2(t_f)\rangle-2\omega^2\text{Re}[\langle \hat{n}_{ph}(t_f)\hat{n}_{ph}(t_i)\rangle].
\eal
Higher moments can be obtained formally from higher orders of the Zassenhaus formula.

\section{LZS interference} 
\label{se:coherence}
We consider $H_{TLS}$ of Eq.~(\ref{eq:TLS}) at $n_g=1/2$. 
Our initial state is $ |\uparrow\rangle\otimes|\sqrt{n_{ph}}\rangle$ and we are interested at the evolved state at specific time $\tau_m=\frac{(2m+1)\pi}{\hbar \omega}$. For this purpose it is convenient to rewrite Eq.~(\ref{eq:TLS}) as
$H_{TLS}=H_0+H_{E_J}$
where
\bea
H_0 &=& H_\uparrow \otimes|\uparrow\rangle\langle \uparrow|+H_\downarrow \otimes|\downarrow\rangle\langle \downarrow|, \nonumber \\
H_{E_J}&=&-\frac{E_J}{2} I\otimes (|\uparrow\rangle\langle \downarrow|+|\downarrow\rangle\langle \uparrow|),
\eea
and the operators $H_{\uparrow,\downarrow}$ act only on the agent,
\bal
H_\uparrow=&\hbar \omega \hat{a}^\dagger \hat{a},\nn\\
H_\downarrow=&\hbar \omega(\hat{a}^\dagger+\frac{\lambda}{\omega})(\hat{a}+\frac{\lambda}{ \omega})-\frac{\hbar \lambda^2}{ \omega}.
\eal

\label{se:perturbation}
Let us expand the evolution operator from $t_i$ to $t_f$, with $t_f-t_i=\tau= \frac{\pi(2m+1)}{\omega}$, up to first order in $E_J$. Defining the unperturbed propagator as $U_0(t_2,t_1)=\exp(-iH_0 (t_2-t_1)/\hbar)$, we have  
\bal
U(\tau,0) &= e^{-i H_{TLS} \tau /\hbar} \simeq U_0(\tau,0)-\frac{i}{\hbar}\int_0^{\tau}dt_1 U_0(\tau,t_1)H_{E_J}U_0(t_1,0)+\cdots.
\eal 
The evolved state becomes
\bal
|\psi(t_f)\rangle=|-\sqrt{n_{ph}}\rangle|\uparrow\rangle  
+\frac{iE_J}{2\hbar} \int_0^{\tau}dt e^{-iH_\downarrow (\tau-t)}|
\sqrt{n_{ph}}e^{-i\omega t}\rangle|\downarrow\rangle +\cdots .
\eal
Up to second order in $E_J$, the probability to find the system in state $|\downarrow \rangle$ (namely the ground state) at $t_f$ is given by
\bea 
P_\downarrow&=&|\langle \downarrow|\psi(t_f)\rangle |^2 = (E_J/2\hbar)^2 \mathcal{I},  \\ 
\mathcal{I}&=& \int_0^\tau dt_1 \int_0^\tau dt_2 \langle \sqrt{n_{ph}} |e^{i H_{\uparrow} t_2/\hbar} e^{i H_{\downarrow} (t_1-t_2)/\hbar} e^{-i H_{\uparrow} t_1/\hbar} | \sqrt{n_{ph}} \rangle. \nonumber
\eea
For each pair $(t_1,t_2)$, this is the overlap between the two paths in which the spin occurs at times $t_1$ and $t_2$, respectively.

To compute the double integral, we define the shifted operator $\hat{a}_\downarrow=\hat{a}+\frac{\lambda}{ \omega}$ and note that its coherent state, satisfying $\hat{a}_\downarrow|\alpha\rangle_\downarrow=\alpha|\alpha\rangle_\downarrow$, is proportional to a coherent state of $\hat{a}$,
\bal
|\alpha\rangle=e^{-i\frac{\lambda}{\omega}\text{Im}[\alpha]}|\alpha+\frac{\lambda}{\omega}\rangle_\downarrow,\label{eq:coherent_relation}
\eal
To see this, we note that the coherent state $|\alpha\rangle$ of $\hat{a}$ must be proportional to $|\alpha+\frac{\lambda}{\omega}\rangle_\downarrow$, but there is an additional phase factor. To obtain the additional phase factor, we compute the overlap $_\downarrow\langle \alpha+\frac{\lambda}{\omega}|\alpha\rangle$. $|\alpha\rangle$ is written as
\bal
|\alpha\rangle=e^{-\frac{|\alpha|^2}{2}}\sum_{k=0}^\infty\frac{\alpha^k}{k!}(\hat{a}^\dagger)^k|0\rangle.
\eal
Using $|0\rangle=|\frac{\lambda}{\omega}\rangle_{\downarrow}$, then
\bal
e^{-\frac{|\alpha|^2}{2}}\sum_{k=0}^\infty\frac{\alpha^k}{k!}(a_\downarrow^\dagger-\frac{\lambda}{\omega})^k|\frac{\lambda}{\omega}\rangle_{\downarrow}.
\eal
Then the overlap is
\bal
 _\downarrow\langle \alpha+\frac{\lambda}{\omega}|\alpha\rangle=&e^{-\frac{|\alpha|^2}{2}}\sum_{k=0}^\infty {}_\downarrow\langle \alpha+\frac{\lambda}{\omega}|\frac{\alpha^k}{k!}(a_\downarrow^\dagger-\frac{\lambda}{\omega})^k|\frac{\lambda}{\omega}\rangle_{\downarrow}=e^{-\frac{|\alpha|^2}{2}}\sum_{k=0}^\infty {}_\downarrow\langle \alpha+\frac{\lambda}{\omega}|\frac{\lambda}{\omega}\rangle_{\downarrow}\frac{|\alpha|^{2k}}{k!} \nn\\
=&e^{\frac{|\alpha|^2}{2}}e^{-\frac{1}{2}|\alpha+\frac{\lambda}{\omega}|^2-\frac{1}{2}\frac{\lambda^2}{\omega^2}+(\alpha^*+\frac{\lambda}{\omega})\frac{\lambda}{\omega}}=e^{-i\frac{\lambda}{\omega}\text{Im}[\alpha]},
\eal
with the desired phase factor in Eq.~(\ref{eq:coherent_relation}).

It follows that
\bal
&e^{-iH_\downarrow (\tau-t)}|\sqrt{n_{ph}}e^{-i\omega t}\rangle|\downarrow\rangle \nn\\
=&e^{-iH_\downarrow (\tau-t)}e^{i\frac{\lambda}{\omega}\sqrt{n_{ph}}\sin(\omega t)}|\sqrt{n_{ph}}e^{-i\omega t}+\frac{\lambda}{\omega}\rangle_\downarrow|\downarrow\rangle \nn\\
=&e^{i\frac{\lambda}{\omega}\sqrt{n_{ph}}\sin(\omega t)}e^{i\frac{\lambda^2}{\omega}(\tau-t)}|(\sqrt{n_{ph}}e^{-i\omega t}+\frac{\lambda}{\omega})e^{-i\omega(\tau-t)}\rangle_\downarrow |\downarrow\rangle, 
\eal
giving
\bal
\mathcal{I}
&=\int_0^{\frac{(2m+1)\pi}{\hbar\omega}} dt_1 dt_2 ~_\downarrow\langle -\sqrt{n_{ph}}-\frac{\lambda}{\omega}e^{i\omega t_2}|-\sqrt{n_{ph}}-\frac{\lambda}{\omega}e^{i\omega t_1}\rangle_\downarrow 
 e^{i\frac{\lambda}{\omega}\sqrt{n_{ph}}(\sin(\omega t_1)-\sin(\omega t_2))}e^{i\frac{\lambda^2}{\hbar\omega}(t_2-t_1)}\nn\\
&=\int_0^{\frac{(2m+1)\pi}{\hbar\omega}} dt_1 dt_2 e^{i\frac{\lambda^2}{\omega}(t_2-t_1)-\frac{\lambda^2}{\omega^2}(1-e^{i\omega(t_1-t_2)})} e^{-\frac{2\lambda\sqrt{n_{ph}}}{\omega}i(\sin(\omega t_2)-\sin(\omega t_1))}\label{eq:dephasing}.
\eal
The resulting Eq.~(\ref{eq:dephasing}) is the starting point of the remaining calculations in this section. While it can be calculated exactly, see Eq.~(\ref{eq:LZS_pert}), we first would like to  understand few limits from this integral. 
\subsection{Classical limit}
First consider the classical limit $\lambda\to 0$ with  $n_{ph} \to \infty$, such that 
  $\lambda\sqrt{n_{ph}}/\omega \to \rm{finite~constant}$, in which $\hat{a}+\hat{a}^\dagger$ describes  a purely external time dependent parameter. Then the double integral in  Eq.~(\ref{eq:dephasing}) becomes the absolute value squared of a single integral, which itself describes the total quantum amplitude,
\bal
\label{eq:saddle_point}
\mathcal{I}_{\substack{\lambda\to 0\\ \frac{\lambda\sqrt{n_{ph}}}{\omega}\to {\rm{finite}}}} =\Bigg|\int_0^{\frac{(2m+1)\pi}{\hbar\omega}}dt_1 e^{i\frac{2\lambda\sqrt{n_{ph}}}{\omega}\sin(\omega t_1)}\Bigg|^2.
\eal
For simplicity, let us assume that $\lambda\sqrt{n_{ph}}$ is very large which allows us to evaluate this integral saddle-point approximation.

First, for $m=0$, we obtain
\bal
\int_0^{\frac{\pi}{\hbar\omega}}dt_1 e^{i\frac{2\lambda\sqrt{n_{ph}}}{\omega}\sin(\omega t_1)}\simeq \int_{-\infty}^{\infty} dt_1 e^{i\frac{2\lambda\sqrt{n_{ph}}}{\omega}(1-\frac{1}{2}\omega^2 t_1^2)}
=e^{i\frac{2\lambda\sqrt{n_{ph}}}{\omega}}\sqrt{\frac{\pi}{i\lambda\sqrt{n_{ph}}\omega}}.
\eal
The phase $e^{i\frac{2\lambda\sqrt{n_{ph}}}{\omega}}$ corresponds to the dynamical phase accumulated from $t=\frac{\pi}{2\hbar \omega}$ to $\frac{\pi}{\hbar \omega}$. This indicates that the spin flip occurs around $t=\frac{\pi}{2 \hbar \omega}$, when the gap is the smallest.

After taking the absolute square of the integral, we obtain the probability $P_\downarrow=\frac{\pi E_J^2}{4\hbar^2\lambda \omega \sqrt{n_{ph}}}$, which coincides with the limit $E_J\to 0$ of the Landau-Zener probability $P_\downarrow=1-e^{-\frac{\pi  E_J^2}{2\hbar^2 \lambda  \omega \sqrt{n_{ph}}}}$. 

Next, consider $m\ne 0$. The $\sin(\omega t_1)$ in the argument of the exponential in Eq.~(\ref{eq:saddle_point}) has $m+1$ maxima and $m$ minima. Treating these extreme via the saddle-point approximation, we have
\bal
\int_0^{\frac{(2m+1)\pi}{\hbar\omega}}dt_1 e^{i\frac{2\lambda\sqrt{n_{ph}}}{\omega}\sin(\omega t_1)}
& \simeq (m+1)\int_{-\infty}^{\infty} dt_1 e^{i\frac{2\lambda\sqrt{n_{ph}}}{\omega}(1-\frac{1}{2}\omega^2 t_1^2)} +m\int_{-\infty}^{\infty} dt_1 e^{-i\frac{2\lambda\sqrt{n_{ph}}}{\omega}(1-\frac{1}{2}\omega^2 t_1^2)}\nn\\
&=(m+1)e^{i\frac{2\lambda\sqrt{n_{ph}}}{\omega}}\sqrt{\frac{\pi}{i\lambda\sqrt{n_{ph}}\omega}}+me^{-i\frac{2\lambda\sqrt{n_{ph}}}{\omega}}\sqrt{\frac{\pi}{-i\lambda\sqrt{n_{ph}}\omega}}.
\eal
Note that the phase $e^{ i\frac{2\lambda\sqrt{n_{ph}}}{\omega}}$ is the dynamical phase accumulated from $t=\frac{\pi}{2\omega}(4n+1)$ ($n=0,1,2,\cdots m$) to $t=\frac{\pi}{\omega}(2m+1)$ and $e^{-i\frac{2\lambda\sqrt{n_{ph}}}{\omega}}$ is the dynamical phase accumulated from $t=\frac{\pi}{2\omega}(4n+3)$ to $t=\frac{\pi}{\omega}(2m+1)$. 
Taking the absolute value squared of this equation, we obtain
\bal
\label{eq:I_classical}
\mathcal{I}=\frac{\pi}{\lambda \sqrt{n_{ph}}\omega}\Big[(2m^2+2m+1)-2m(m+1)\sin \frac{4\lambda\sqrt{n_{ph}}}{\omega}\Big].
\eal
We can inspect oscillations versus $n_{ph}$, due to dynamical phase, however there is no oscillation versus $m$. This result is plotted in Fig.~\ref{fig:2d_plot}, and compared with the full simulation.

\subsection{Agent driving}
Now we compute Eq.~(\ref{eq:dephasing}) for finite $\lambda$ and finite $n_{ph}$, and study the deviation from the classical limit. To compute the integral, we expand it as a power series in $\lambda/\omega$,
\bal
\label{eq:sum_square_int}
\mathcal{I}= \sum_{n=0}^\infty \frac{1}{n!}\frac{\lambda^{2n}}{ \omega^{2n}}e^{-\frac{\lambda^2}{\omega^2}}\int_0^{\frac{(2m+1)\pi}{\omega}} dt_1 dt_2 e^{i\frac{\lambda^2}{\omega}(t_2-t_1)}e^{in\omega(t_1-t_2)}  e^{-\frac{2\sqrt{n_{ph}}\lambda}{ \omega}i(\sin(\omega t_2)-\sin(\omega t_1))}.
\eal
The double integral in Eq.~(\ref{eq:dephasing}) is now a sum over a products of two independent integrals of the form
\bal
\int_0^{(2m+1)\frac{\pi}{\hbar\omega}} dt_1 e^{i(n\omega-\frac{\lambda^2}{\omega}) t_1+i\frac{2\lambda\sqrt{n_{ph}}}{\omega}\sin(\omega t_1)}\int_0^{(2m+1)\frac{\pi}{\omega}}dt_2 e^{-i(n\omega-\frac{\lambda^2}{\omega}) t_2-i\frac{2\lambda\sqrt{n_{ph}}}{\omega}\sin(\omega t_2)}.
\eal
Before performing the integral explicitly, let is again apply the saddle-point approximation by assuming that $\lambda\sqrt{n_{ph}}$ is large. The first integrand can be written as $e^{f(t)}$ with
\bal
f(t_1)=&i(n\omega-\frac{\lambda^2}{\omega}) t_1+i\frac{2\lambda\sqrt{n_{ph}}}{\omega}\sin(\omega t_1),\nn\\
\partial_{t_1}f(t_1)=&i(n\omega-\frac{\lambda^2}{\omega}) +i2\lambda\sqrt{n_{ph}}\cos(\omega t_1)=0.
\eal
Now $f(t_1)$ has extrema at approximately $t_{1,0}\simeq \frac{\pi(4p+1)}{2\omega}+\frac{n-\frac{\lambda^2}{(\hbar \omega)^2}}{2\lambda\sqrt{n_{ph}}}$ and $\frac{\pi(4q+3)}{2\omega}-\frac{n-\frac{\lambda^2}{(\hbar\omega)^2}}{2\lambda\sqrt{n_{ph}}}$, where $p$ ($q$) is an integer number smaller than $m$ ($m-1$). These points correspond to the avoided crossings, up to small corrections $\propto \frac{1}{\sqrt{n_{ph}}}$, which signal small deviations from the Born-Oppenheimer condition. In this case the trajectory of the agent is weakly affected by the spin-flips occurring in the system.
Then the last integral becomes
\bal
\int_0^{(2m+1)\frac{\pi}{\hbar\omega}}dt_1 e^{i(n\omega-\frac{\lambda^2}{\omega}) t_1+i\frac{2\lambda\sqrt{n_{ph}}}{\hbar\omega}\sin(\omega t_1)}
\simeq &\sum_{p=0}^{m}e^{i(n\omega-\frac{\lambda^2}{\hbar^2\omega})(\frac{\pi(4p+1)}{2\omega}+\frac{n-\frac{\lambda^2}{\omega^2}}{2\lambda\sqrt{n_{ph}}})}e^{i\frac{2\lambda\sqrt{n_{ph}}}{\omega}} e^{-i\frac{1}{2} \omega\frac{(n-\frac{\lambda^2}{\omega^2})^2}{2\lambda\sqrt{n_{ph}}}}\int_{-\infty}^{\infty}d\delta t_1 e^{-i\frac{1}{2}2\lambda\sqrt{n_{ph}}\omega\delta t_1^2}\nn\\
&+\sum_{q=0}^{m-1}e^{i(n \omega-\frac{\lambda^2}{\omega})(\frac{\pi(4q+3)}{2\omega}-\frac{n-\frac{\lambda^2}{(\hbar\omega)^2}}{2\lambda\sqrt{n_{ph}}})}e^{-i\frac{2\lambda\sqrt{n_{ph}}}{\hbar\omega}} e^{i\frac{1}{2}\omega\frac{(n-\frac{\lambda^2}{ \omega^2})^2}{2\lambda\sqrt{n_{ph}}}}\int_{-\infty}^{\infty}d\delta t_1 e^{i\frac{1}{2}2\lambda\sqrt{n_{ph}}\omega\delta t_1^2}.
\eal
Skipping the calculation of the Gaussian integral and summation over $m$, we finally obtain
\bal
\mathcal{I}\simeq & \frac{\pi e^{-\frac{\lambda^2}{\omega^2}}}{\lambda\omega\sqrt{n_{ph}}}\sum_{n=0}^{\infty}\Big[\frac{\sin^2 (\pi (m+1) (n-\frac{\lambda ^2}{\omega ^2}))}{\sin^2 (\pi  (n-\frac{\lambda ^2}{\omega ^2}))}+\frac{\sin^2 (\pi  m (n-\frac{\lambda ^2}{\omega ^2}))}{\sin^2 (\pi  (n-\frac{\lambda ^2}{\omega ^2}))} \nonumber \\
 & ~~~~~~~~~~~~~~~~~~~~~~~~~~~~~~~~~+2\sin(\frac{\lambda\sqrt{n_{ph}}}{\omega}+\omega\frac{(n-\frac{\lambda^2}{\omega^2})^2}{2\lambda\sqrt{n_{ph}}})\frac{\sin(\pi(m+1)(n-\frac{\lambda^2}{\omega^2}))\sin (\pi  m (n-\frac{\lambda ^2}{\omega ^2}))}{\sin^2 (\pi  (n-\frac{\lambda ^2}{\omega ^2}))}\Big]\nn\\
\simeq &\frac{\pi }{2\lambda\omega\sqrt{n_{ph}} \sin ^2\left(\frac{\pi  \lambda ^2}{\omega ^2}\right)}\Bigg[2 e^{-\frac{2 \lambda ^2}{\omega ^2}} \sin \left(\frac{\sqrt{n_{ph}} \lambda }{\omega }\right)\left(\cos \left(\frac{\pi  \lambda ^2}{\omega ^2}\right)-\cos \left(\frac{\pi  \lambda ^2 (2 m+1)}{\omega ^2}\right)\right)\nonumber \\
 & ~~~~~~~~~~~~~~~~~~~~~~~~~~~~~~~~~+2-\cos \left(\frac{2 \pi  \lambda ^2 (m+1)}{\omega ^2}\right)-\cos \left(\frac{2 \pi  \lambda ^2 m}{\omega ^2}\right)\Bigg].
\eal
As opposed to the classical driving case Eq.~(\ref{eq:I_classical}), here there are oscillations versus $m$. As far as the goal of reproducing the classical driving using the agent is concerned, these oscillations are an artifact of the correction to the dynamical phase coming from the deviations from the Born-Oppenheimer condition. We can see that the relative size of these corrections is  $\mathcal{O}\left(\frac{\lambda^2}{ \omega^2} \right)$.

\subsection{Exact evaluation of (\ref{eq:dephasing})}
Using the expansion in Eq.~(\ref{eq:sum_square_int}), the integral is computed as
\bal
\mathcal{I}=\frac{\pi^2}{\omega^2} \Big|\frac{\sin(\pi(n-\frac{\lambda^2}{\omega^2})(m+1))}{\sin(\pi(n-\frac{\lambda^2}{\omega^2}))}\Phi_{-n+\frac{\lambda^2}{\omega^2}}^*(x)+\frac{\sin(\pi(n-\frac{\lambda^2}{\omega^2})m)}{\sin(\pi(n-\frac{\lambda^2}{\omega^2}))} \Phi_{n-\frac{\lambda^2}{\omega^2}}(x)\Big|^2,
\eal
where $x=2\lambda\sqrt{n_{ph}}/\omega$ and $\Phi_{\nu}(x)=J_\nu(x)+i E_\nu(x)$, where $J_\nu(x)$ is the Anger function and $E_\nu(x)$ is the Weber function. Then the integral in Eq.~(\ref{eq:dephasing}) becomes
\bal
\mathcal{I}= e^{-\frac{\lambda^2}{\omega^2}} \sum_{n=0}^\infty \frac{1}{n!}\frac{\lambda^{2n}}{\omega^{2n}}\frac{\pi^2}{\omega^2} \Big|\frac{\sin(\pi(n-\frac{\lambda^2}{\omega^2})(m+1))}{\sin(\pi(n-\frac{\lambda^2}{\omega^2}))}\Phi_{-n+\frac{\lambda^2}{\omega^2}}^*(x)+\frac{\sin(\pi(n-\frac{\lambda^2}{\omega^2})m)}{\sin(\pi(n-\frac{\lambda^2}{\omega^2}))} \Phi_{n-\frac{\lambda^2}{\omega^2}}(x)\Big|^2.\label{eq:LZS_pert}
\eal
As explained above, note that the probability contains oscillations both versus $m$ and $n_{ph}$. In the limit of classical driving $(\lambda\to 0)$, the oscillation of $m$ disappears and becomes a quadratic function, see Eq.~(\ref{eq:I_classical}). This perturbative expression in $E_J$ nicely compares with the simulation of the full model as displayed in  Fig.~\ref{fig:2d_plot111}.

\begin{figure}
    \centering
    \includegraphics[width=0.5\linewidth]{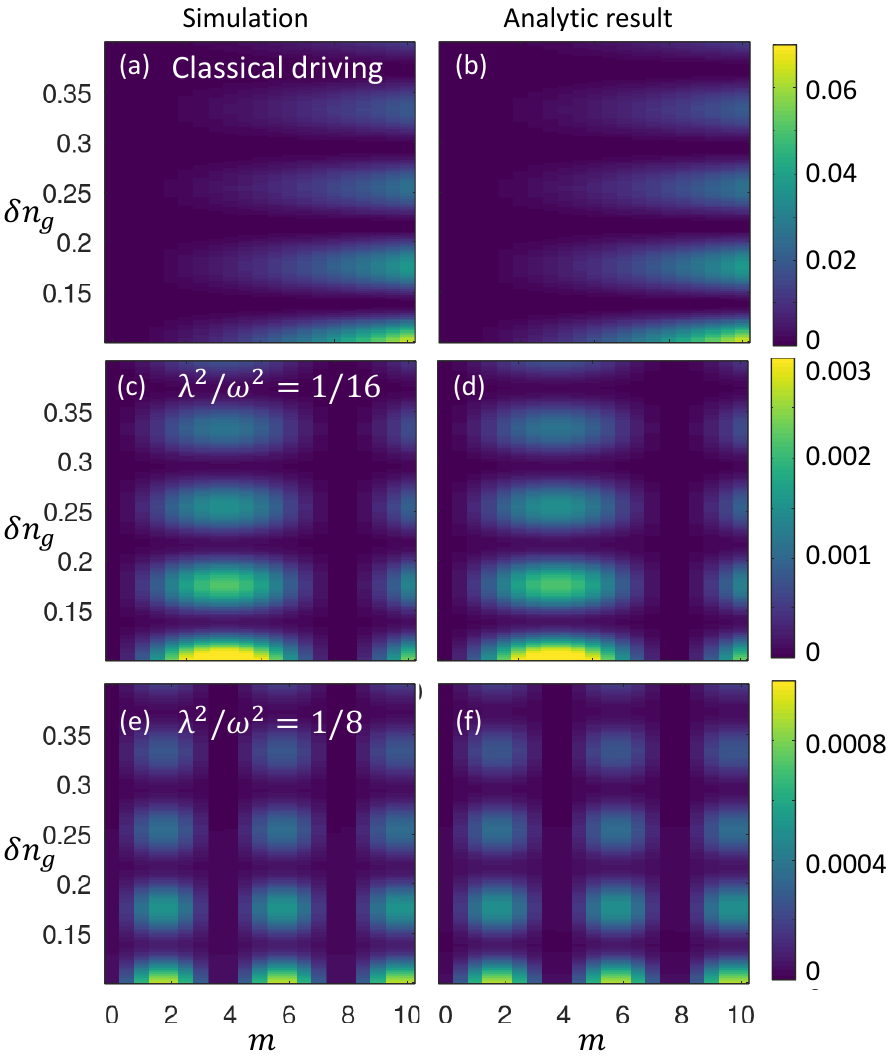}
    \caption{Transition probability $P_\downarrow$ for various $\lambda/ \omega$. Here we set $E_c=\hbar\omega$ and $E_J=0.01\hbar \omega$, where the perturbative calculation is applicable. The plots in the right column are obtained from Eq.~(\ref{eq:LZS_pert}).  }
    \label{fig:2d_plot111}
\end{figure}


\section{Relation to decay and revivals of Rabi oscillations}
\label{appendix:JC}
 The ground and excited states of $H_{TLS}$ at  $n_g=1/2$   are $|g\rangle = (|\uparrow \rangle - |\downarrow \rangle)/\sqrt{2}$ and $|e\rangle = (|\uparrow \rangle + |\downarrow \rangle)/\sqrt{2}$. In this basis, $\sigma_z=(|g\rangle \langle e| +|e\rangle \langle g| )$, and
\bea
H_{TLS}&=&\hbar \omega \hat{a}^\dagger \hat{a} + \frac{E_J}{2}(|e\rangle \langle e| -|g\rangle \langle e| ) ) + \lambda \frac{ \hat{a} +\hat{a}^\dagger}{2} +[-\frac{\lambda}{2}( \hat{a} +\hat{a}^\dagger)  +4E_c(n_g-1/2)](|g\rangle \langle e| +|e\rangle \langle g| ).
\eea
We can absorb the $\lambda \frac{ \hat{a} +\hat{a}^\dagger}{2}$ term by a redefinition of the creation and annihilation operators, $\hat{a}'=\hat{a}+\frac{\lambda}{2 \hbar \omega}$. In terms of these new operators, selecting $n_g$ such that
\be
\frac{\lambda^2}{2 \hbar \omega} + 4E_c(n_g-1/2)=0,
\ee
we obtain
\bea
H_{TLS}=\hbar \omega {\hat{a}}'^\dagger {\hat{a}}'+ \frac{E_J}{2}(|e\rangle \langle e| -|g\rangle \langle e| ) ) -\frac{\lambda}{2}( {\hat{a}}'+{\hat{a}}'^\dagger)  (|g\rangle \langle e| +|e\rangle \langle g| ).
\eea
Let us consider the resonant limit $\hbar \omega \approx E_J$. In this case absorption of a photon exactly gives the excitation energy from $|g\rangle$ to $|e\rangle$. In this case, when the Rabi frequency is small compared to the excitation energy, $\lambda \ll \hbar \omega \approx  E_J$,  it is legitimate to neglect the counter rotating terms and obtain the Jaynes–Cummings Hamiltonian,
\bea
H_{TLS}=\hbar \omega \hat{a}'^\dagger \hat{a}' + \frac{E_J}{2}(|e\rangle \langle e| -|g\rangle \langle g| ) ) -\frac{\lambda}{2} (\hat{a}'^\dagger|g\rangle \langle e| +\hat{a}'|e\rangle \langle g| ).
\eea
The Rabi oscillations occurring when the creation and annihilation operators are replaced by a classical field, are strongly modified when one considers a mesoscopic coherent state with a not-too-large value of $n_{ph}$. This is the model studied in~\cite{meunier2005rabi}

With an initial state $|g\rangle \otimes |\alpha \rangle$ with $|\alpha \rangle = \sum_{n=0}^\infty c_n |n\rangle$ with $c_n = \exp( -|\alpha|^2/2)  \alpha^n/\sqrt{n!}$, let us follow~\cite{meunier2005rabi} and compute the probability to find the excited state at time $t$. Let us set the resonant condition $\hbar \omega=E_J$. Then the eigenstates are $|g,0 \rangle$ and
\bea
|\pm \rangle_n=(|g,n\rangle \pm |e,n-1\rangle)/\sqrt{2}.
\eea
We can expand the initial state in eigenstates,
\be
|\Psi(0) \rangle =|g\rangle \otimes |\alpha \rangle = c_0 |g,0 \rangle +\sum_{n=1}^\infty c_n \frac{|+ \rangle_n+|- \rangle_n}{\sqrt{2}}.
\ee
Therefore
\be
\langle e| \Psi(t) \rangle = \sum_{n=1}^\infty c_n \frac{e^{-i E_{|+ \rangle_n} t/\hbar}-e^{-i E_{|- \rangle_n} t/\hbar}}{2},
\ee
hence the time dependent excitation probability
is given by~\cite{meunier2005rabi}
\be
\label{eq:Haroche}
P_e(t) = \sum_{n=1}^\infty c_n \sin^2 \frac{(E_{|+ \rangle_n}-E_{|- \rangle_n} )t}{2\hbar}=\sum_{n=1}^\infty c_n \sin^2 \frac{\lambda \sqrt{n} t}{2\hbar},
\ee
which displays collapse and revivals as seen in Fig.~\ref{fig:collapse_revival}.
Suppose first that we replace $n$ by its average $\langle n \rangle =\sum_{n=0}^\infty c_n^2 n=n_{ph}$. The resulting Rabi period $T_R$ is defined by $\pi= \frac{\lambda \sqrt{n_{ph}} T_R}{2 }$. In Fig.~\ref{fig:collapse_revival} we plot $P_e(t)$ in these units for $n_{ph}=13.4$ as in ~\cite{meunier2005rabi}. We see first a collapse of the oscillations due to the dispersion of frequencies. The dispersion of frequencies turns out to be of order $\lambda$. Therefore $t_{collapse}  =1/\lambda$. So there are $\sim \sqrt{n_{ph}}$ oscillation before the decay~\cite{meunier2005rabi}.   Interestingly, then there is a revival. By setting the relative phase of two neighboring frequencies to be $2\pi$,
\be
2\pi = \frac{\lambda \sqrt{n+1} t_{revival}}{2} - \frac{\lambda \sqrt{n} t_{revival}}{2},
\ee
we obtain
\be
t_{revival} = \frac{\pi  \sqrt{ n_{ph}}}{\lambda}. 
\ee
So there are $\sim n_{ph}$ oscillations till recovery~\cite{meunier2005rabi}.
\begin{figure}
    \centering
    \includegraphics[width=0.4\linewidth]{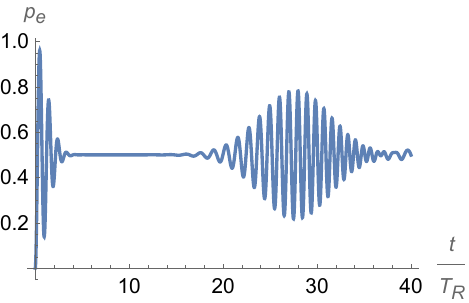}
    \caption{Collapse and revival of Rabi oscillations according to Eq.~(\ref{eq:Haroche}).} 
    \label{fig:collapse_revival}
\end{figure}
\end{widetext}
\end{appendix}


\bibliography{bibliogaphy.bib}
\end{document}